\documentclass[a4paper,twoside]{article}
\usepackage{AMS2LA,fancyhdr,CJK,xcolor,multicol,graphics,greekup,bm,cmap}
\usepackage{amssymb}\usepackage{gensymb,mathcomp}
\usepackage[T1]{fontenc}
\usepackage{cite}
\usepackage{graphicx}
\usepackage{float}
\usepackage[numbers,sort&compress]{natbib}
\usepackage[pdfstartview=FitH, bookmarks=false, colorlinks=false,
    linkcolor=blue, urlcolor=blue, pdfborder=001, citecolor=blue,
    pdftitle={Changepdftitle}, pdfauthor={Changepdfauthor},
    pdfcreator=CHIN. PHYS. LETT., pdfproducer=CPL,
    pdfkeywords={changePACS}]{hyperref}
\newcommand{\explel}{{\large\sffamily~~~~~~~~~~~~~~~~~} \hfill}
\newcommand{\expler}{\hfill\fcolorbox{lightgray}{lightgray}{\textcolor{red}{{\large\fontfamily{phv}\textbf{New Submission}}}}}

\usepackage{color}
\definecolor{dgreen}{cmyk}{1.,0.,1.,0.2}        
\definecolor{orange}{cmyk}{0.,0.353,1.,0.}    

\def\headrule{\kern 1mm \hrule width 17cm \kern -1mm}%
\def\footnoterule{\kern 1mm \hrule width 7cm \kern 2.2mm}%
\newcommand{\cplyear}{2021} \newcommand{\cplvol}{x}
\newcommand{\cplno}{x} \newcommand{\cplpagenumber}{xx{xxxx}}
 \newcommand{\cplpage}{\cplpagenumber-\thepage}

\begin{document}
\begin{CJK*}{GBK}{song}\vspace* {-6mm} \begin{center}
\large\bf{\boldmath{Dynamical exploring the QCD matter at finite temperatures and densities-a short review}}
\footnotetext{\hspace*{-5.4mm}

\noindent$^{*}$Corresponding author. Email: huichaosong@pku.edu.cn


\noindent\copyright\,{\cplyear}
\href{http://www.cps-net.org.cn}{Chinese Physical Society} and
\href{http://www.iop.org}{IOP Publishing Ltd}}
\\[6mm]
\normalsize \rm{}Shanjin Wu$^{1,2,3}$, Chun Shen$^{4, 5}$, and Huichao Song$^{2,3,1*}$
\\[2mm]\small\sl $^{1}$Center  for  High  Energy  Physics,  Peking  University,  Beijing  100871,  China

$^{2}$Department of Physics and State Key Laboratory of Nuclear Physics and Technology, Peking University, Beijing 100871, China

$^{3}$Collaborative Innovation Center of Quantum Matter, Beijing 100871, China

$^{4}$ Department of Physics and Astronomy, Wayne State University Detroit, Michigan 48201, USA

$^{5}$ RIKEN BNL Research Center, Brookhaven National Laboratory Upton, New York 11973, USA
\\[4mm]\normalsize\rm{}(Received x April 2021; accepted xxx; published online )
\end{center}
\end{CJK*}
\vskip 1.5mm

\noindent{\narrower\small{}\textbf{Abstract}
We provide a concise review on recent theory advancements towards full-fledged (3+1)D dynamical descriptions of relativistic nuclear collisions at finite baryon density. Heavy-ion collisions at different collision energies produce strongly-coupled matter and probe the QCD phase transition at the crossover, critical point, and first-order phase
transition regions. Dynamical frameworks provide a quantitative tool to extract properties of hot QCD matter and map fireballs to the QCD phase diagram.  Outstanding challenges are highlighted when confronting current theoretical frameworks with current and forthcoming experimental measurements from the RHIC beam energy scan programs.
\par}\vskip 3mm
\normalsize\noindent{\narrower{PACS:  12.38.Mh,21.65.Qr,25.75.-q,64.60.-i}
{\rm\hspace*{13mm}DOI: 10.1088/0256-307X/\cplvol/\cplno/\cplpagenumber}

\par}\vskip 6mm
\begin{multicols}{2}

\section{Introduction}

The phase transition and phase structure of the strongly interacting matter at finite temperature and density are central topics in high-energy nuclear physics. Lattice QCD simulations show that the phase transition at vanishing baryon chemical potential ($\mu_B\simeq0$) is a smooth crossover \cite{Aoki:2006we,Ding:2015ona,Bazavov:2019lgz,Ratti:2018ksb}. Effective field theory models predict a first-order phase transition boundary at large chemical potential, together with a critical endpoint~\cite{Fukushima:2010bq,Fukushima:2013rx,Fischer:2018sdj}. However, lattice QCD suffers from the notorious sign problem at large chemical potential region, and theoretical predictions for the critical point's location remain model-dependent~\cite{Stephanov:2004wx,Stephanov:2007fk}.

On the experimental side, relativistic heavy-ion collisions at Relativistic Heavy-Ion Collider (RHIC) and the Large Hadron Collider (LHC) aim to create the quark-gluon plasma (QGP) and explore its phase transition at zero and finite baryon chemical potential. At top RHIC and the LHC energies, the created QGP behaves like a nearly perfect liquid with almost vanishing chemical potential at mid-rapidity~\cite{Gyulassy:2004vg,Gyulassy:2004zy,Kolb:2003dz,Muller:2012zq}. To probe the QCD phase diagram at finite baryon chemical potential and search the QCD critical point, RHIC has carried out Beam Energy Scan (BES) program for Au+Au collisions with collision energies ranges from $3 \sim 200$ A GeV~\cite{Aggarwal:2010cw,Luo:2017faz,Bzdak:2019pkr}.
Future experimental programs, such as Facility for Antiproton and Ion Research (FAIR) in Darmstadt \cite{Friman:2011zz}, Nuclotron-based Ion Collider fAcility (NICA) in Dubna \cite{NICA-web}, HIAF in Huizhou
\cite{Ruan:2018fpo} and J-PARC-HI in Tokai \cite{J_PARC-HI-web}, will further explore the phase diagram
at higher baryon density to search for the QCD critical point, the first-order phase transition boundary, and study the strongly interacting matter at high baryon density.

On the theoretical and phenomenological side, significant progress has been achieved to extract the QGP transport properties at top RHIC and LHC energies~\cite{Song:2010mg,Song:2012ua,Bernhard:2016tnd,
Bernhard:2019bmu,Nijs:2020roc,Everett:2020yty,Everett:2020xug}.
In particular, integrated hybrid models have been developed to quantitatively describe the complex multi-stage evolution of the QCD matter created in relativistic heavy-ion collisions \cite{Song:2010mg,Song:2010aq,Hirano:2012kj,Shen:2014vra,Niemi:2015qia,Ryu:2017qzn,Pang:2018zzo,Shen:2017bsr}. To extend the theoretical frameworks to heavy-ion collisions at $\sqrt{s_\mathrm{NN}} \sim \mathcal{O}(10)$\,GeV, new theory ingredients in multiple aspects need to be included. In particular, the QCD phase transition must be properly encoded in the dynamical model to describe the experimental measurements quantitatively. Phenomenological studies with upcoming precise measurements can extract robust QGP transport coefficients in a baryon-rich environment.

This short review will summarize recent progress in improving the dynamical model descriptions of relativistic nuclear collisions as they go through crossover, critical point, and first-order phase transition regions in the QCD phase diagram.

\section{Quantitative characterization of the QCD crossover region}
In this section, we focus on dynamical descriptions of relativistic heavy-ion collisions with collision energies $\sqrt{s_\mathrm{NN}} \ge 20$\,GeV. These collisions probe the hot QCD matter with net baryon chemical potential $\mu_B \le 250$\,MeV \cite{Andronic:2017pug, Adamczyk:2017iwn}, where the QGP and hadronic phases are connected with a smooth crossover.

\underline{\textbf{Hydrodynamics with dynamical initialization}}
\underline{\textbf{schemes:}} Hydrodynamics and hybrid models are important tools to describe the QGP fireball evolution and study soft observables for relativistic heavy-ion collisions at RHIC and LHC energies \cite{Heinz:2013th,Gale:2013da,Jeon:2015dfa,Song:2017wtw}.For collision energies $\sqrt{s_\mathrm{NN}} = 20 - 60$\,GeV, the duration for two colliding nuclei with radii $R$ to pass through each other can be estimated by $\tau_\mathrm{overlap} \sim 2R / \sqrt{\left(\frac{\sqrt{s_\mathrm{NN}}}{2m_N}\right)^2 - 1}$, which gradually increases with the decrease of collision energy~\cite{Karpenko:2015xea, Shen:2017bsr, Shen:2017fnn}. Meanwhile, the longitudinal boost invariance
violates significantly~\cite{Noronha:2018atu, Shen:2020jwv}. Therefore, it is important to study the role of pre-equilibrium dynamics during the overlapping period within the framework of (3+1)D hydrodynamics. In recent years, dynamical initialization schemes have been developed to model this extended interaction region in heavy-ion collisions \cite{Okai:2017ofp, Shen:2017ruz, Shen:2017bsr,Du:2018mpf, Akamatsu:2018olk, Kanakubo:2019ogh}. Commonly, they interweave the initial collision stage with hydrodynamics on a local basis while the two nuclei pass through each other. The initial state energy-momentum and conserved charge density currents are treated as sources to feed the hydrodynamic fields,
\begin{eqnarray}
\partial_\mu T^{\mu\nu} &=& J^\nu_\mathrm{source}(\tau, \vec{x}) \\
\partial_\mu J^\mu_i &=& \rho_{i,\mathrm{source}}(\tau, \vec{x}).
\end{eqnarray}
Here the conserved quantum charges for light flavor quarks are baryon, strangeness, and electric charges, $i = B, S, Q$. Dynamical initialization schemes require initial state models to provide the (3+1)D space-time and momentum information of the energy-momentum and charge distributions. There has been a stream of collective effort to develop 3D initial state models. The complex 3D collision dynamics can be approximated by parametric energy and charge depositions \cite{Hirano:2005xf, Bozek:2010vz, Bozek:2015bna, Bozek:2017qir, Shen:2020jwv, Sakai:2020pjw}. More dynamical models involve simulating energy loss during individual nucleon-nucleon collisions. Such initial state models have been built based on classical string deceleration \cite{Shen:2017bsr, Bialas:2016epd}. And there are 3D initial conditions based on hadronic and partonic transport simulations \cite{Pang:2012he, Karpenko:2015xea,Du:2018mpf,Xu:2016hmp,Fu:2020oxj}. These models provide non-trivial correlations between the longitudinal energy distribution and flow velocity.

Furthermore, recent theory developments to understand early-stage baryon stopping from the Color Glass Condensate-based approaches in the fragmentation region \cite{Li:2018ini, McLerran:2018avb}. The initial energy density and baryon charge distributions were also studied from a holographic approach at intermediate couplings \cite{Attems:2018gou}. Measurements of the rapidity-dependent particle production and flow correlations can provide valuable constraints on initial state longitudinal fluctuations and baryon stopping.

\underline{\textbf{Equation of State (EoS):}}
The equation of state for nuclear matter for $\mu_B \le 250$\,MeV has been constructed based on the Taylor series technique with high-order susceptibility coefficients computed from lattice QCD calculations~\cite{Ratti:2018ksb, Bazavov:2018mes, Monnai:2019hkn, Noronha-Hostler:2019ayj, Parotto:2018pwx}. See a recent review for more details on this topic~\cite{Monnai:2021kgu}. These equations of state at finite densities are essential to enable dynamics of Au+Au collisions at the RHIC BES energies. Furthermore, one needs to build in constraints on the strangeness neutrality and electric charge $n_Q \simeq 0.4 n_B$ \cite{Monnai:2019hkn, Noronha-Hostler:2019ayj} for heavy-ion collisions with gold and lead nuclei. The strangeness neutrality condition was crucial to reproduce the multi-strangeness baryon yields and could significantly change the fireball phase trajectories in the QCD phase diagram \cite{Monnai:2019hkn}.

\underline{\textbf{Transport coefficients:}} 
Transport coefficients are important inputs in hydrodynamic simulations. At the top RHIC and LHC energies, the Bayesian inference method has been adopted to quantitatively constrain the shear and bulk viscosity and their temperature dependence with the soft hadron measurements~\cite{Bernhard:2016tnd,
Bernhard:2019bmu,Nijs:2020roc,Everett:2020yty,Everett:2020xug}. Extending this approach to analyze flow measurements at the RHIC Beam Energy Scan program would further extract the $\mu_B$-dependence of the specific QGP viscosity with reliable uncertainty. Studies have shown that the collision energy and rapidity dependence of the anisotropic flow coefficients has strong constraining power of QGP's $\eta/s(T, \mu_B)$ \cite{Karpenko:2015xea, Denicol:2015nhu, Shen:2020jwv}. To take advantage of those measurements in the Bayesian analysis, full-scale (3+1)D hydrodynamic simulations are essential.

The dynamics of conserved charge currents allow us to study new transport properties of the QGP, namely the charge diffusion processes inside the fluid. Driven by the local gradients of $\mu_i/T$, the net baryon, strangeness, and electric charges can flow with different velocities than the energy density. Recent works showed that the net baryon diffusion has important effects on the longitudinal dynamics of the net baryon current \cite{Denicol:2018wdp, Li:2018fow, Du:2019obx, Wu:2021ypv}. A theoretical framework that incorporates diffusion effects on multiple conserved charge currents has been developed \cite{Fotakis:2019nbq}. Such a framework can help us access the full diffusion coefficient matrix, which takes the cross charge correlation into account \cite{Greif:2017byw, Rose:2020sjv}.

\underline{\textbf{Hadronic transport:}} Hydrodynamics are generally connected with a hadronic transport model for a better description of the non-equilibrium late hadronic evolution~\cite{Song:2010aq,Hirano:2012kj,Shen:2014vra,Ryu:2017qzn}. As the QGP fireball evolves into the phase-transition region, individual fluid cells are converted to hadrons according to the Cooper-Frye prescription \cite{Cooper:1974mv}. These hadrons can further scatter with each other and decay to stable states described by hadronic transport model models, such as UrQMD \cite{Bass:1998ca, Bleicher:1999xi}, JAM \cite{Nara:1999dz}, and SMASH \cite{Weil:2016zrk}. As the collision energy reduces from 200 GeV to 20 GeV, the role of the late-stage hadronic transport becomes more and more crucial \cite{Monnai:2019hkn}. Because the hydrodynamic phase in heavy-ion collisions at $\sqrt{s} \sim 20-60$\,GeV is shorter than those at TeV collision energies at the LHC, both radial and anisotropic flow further develop during the hadronic evolution due to the remaining spatial inhomogeneity~\cite{Shen:2011zc, Denicol:2018wdp}.


\section{Dynamic models near the critical point}
This section will briefly review critical phenomena in equilibrium and then highlight the dynamical descriptions near the QCD critical point.

\subsection{Equilibrium critical phenomena }

One of the landmark features for an equilibrium system near the critical point is the long-range correlation, which results in many unique properties, such as critical opalescence, universal scaling, singular behavior of the equation of state, large fluctuations of thermodynamics variables, and so on. This subsection will briefly review the critical phenomena for an equilibrium hot QCD system.

\underline{\textbf{ Universal Scaling of the hot QCD system}}: Due to the infinite long-range correlation and divergence of the correlation length, the system at the critical point has no characteristic scale,
which behaviors self-similarly under the scale transformation. As a result, different systems in the same universality class, determined by the system's dimension and the number of order parameter components, share universal critical properties. From the symmetry analysis, it is generally believed that QCD critical point and 3D-Ising model belong to the same static universality class~\cite{Pisarski:1983ms,Wilczek:1992sf,Rajagopal:1992qz}. For finite volume systems, such as the QGP fireball created in heavy-ion collisions, the universal scaling argument should take finite size scaling into account~\cite{Palhares:2009tf,Fraga:2011hi}.

\underline{\textbf{EoS with a critical point}}: Recent lattice QCD calculations have narrowed the location of the critical point, which may exist in the region of $T<140,\mu_B>300$ MeV \cite{Ding:2020rtq}. However, it is still challenging for lattice simulations to precisely predict the position of the critical point due to the sign problem at finite $\mu_B$. On the other hand, because the hot QCD system belongs to the same universality class of the 3D-Ising model~\cite{Rajagopal:1992qz,Berges:1998rc,Halasz:1998qr,Karsch:2001nf}, non-universally map the EoS between these two systems~\cite{Nonaka:2004pg,Bluhm:2006av,Parotto:2018pwx,Pradeep:2019ccv} provide a proper parametrization for EoS with a QCD critical point. Recently, a more comprehensive EoS near the critical point has been constructed, which combines the singular part from the Ising model and the background part obtained from lattice EoS at $\mu_B=0$~\cite{Parotto:2018pwx,Mroczek:2020rpm,Stafford:2021wik}. It is crucial to study such EoS effects on final observables, which has not been done yet.


\underline{\textbf{ Multiplicity fluctuations}}: The strong fluctuations of the order parameter field near the critical point could lead to large fluctuations of final hadrons produced in heavy-ion collisions~\cite{Stephanov:1998dy,Stephanov:1999zu}, especially for the event-by-event net-proton fluctuations~\cite{Hatta:2003wn,Kitazawa:2012at,Kitazawa:2011wh}. However, the correlation length is largely limited due to the finite size and finite time effects of the QGP fireball created in heavy-ion collisions~\cite{Berdnikov:1999ph,Nonaka:2004pg}, which make it hard to probe the critical point from the second-order cumulant of the event-by-event multiplicity fluctuations. Ref.~\cite{Stephanov:2008qz} proposed that the higher-order cumulants are more sensitive to correlation length, some of which change signs near the QCD phase boundary~\cite{Stephanov:2011pb,Athanasiou:2010kw,Asakawa:2009aj}. Recently, the higher-order cumulants of the net-proton have been systematically measured in the RHIC BES program. The kurtosis of net-proton in the most central collisions presents non-monotonic behavior, which qualitatively agrees with the theoretical expectation and indicating the potential discovery of QCD critical point. However, the sign of skewness data fails to describe the experiment measurement~\cite{Jiang:2015hri} and requires to include the dynamical effect, which will be discussed in the following subsection.

\subsection{Dynamical fluctuations near the QCD critical point}
For systems evolving near the critical point, the dynamical universality class is classified according to the symmetry of order parameter, dimensionality, conservation laws of the conserved densities, and the Poisson bracket among them~\cite{Hohenberg:1977ym}. Ref.~\cite{Son:2004iv} suggested that the evolving hot QCD system is in the class of model H, which describes a dynamical system with the conserved order parameter, conserved momentum density, and the Poisson bracket between them.
Owing to the complexity of numerical implementations of model H, model A and B have been served as simplified dynamical models near the QCD critical point, which will be briefly reviewed as follow. We will also review the recent progress on other dynamical models near the critical point, such as Non-equilibrium chiral fluid dynamics and the Hydro+ formalism.

\underline{\textbf{ Model A}} only evolves the non-conserved order parameter field $\sigma$, {which serves as a simplified version of the dynamic model near the QCD critical point}. It was found that the critical slowing down effects not only limit the growth of the correlation length, as predicted in early papers~\cite{Nonaka:2004pg,Berdnikov:1999ph}, but also substantially modifies the temporal evolution of cumulants and even reverse the signs of skewness and kurtosis compared to the equilibrium values~\cite{Mukherjee:2015swa,Mukherjee:2016kyu,Jiang:2017mji,Wu:2018twy}.

\underline{\textbf{ Model B}} only evolves the conserved quantity, which can be considered as another simplified model for the evolving dynamical systems near the QCD critical point. It is motivated by the fact that the dynamics of the non-conserved order parameter field becomes insignificant by mixing with the conserved baryon number density, which makes the dynamics of conserved quantity dominate at a long time scale \cite{Son:2004iv}. Besides the normal critical slowing down effects \cite{Nahrgang:2018afz,Sakaida:2017rtj,Wu:2019qfz}, one of the intriguing results for the conserved critical dynamics is the competition between the growth of correlation length and the diffusion effect, which leads to a non-monotonic behavior for the multiplicity fluctuations of the conserved charges as the rapidity window increased \cite{Sakaida:2017rtj}. For more realistic implementation, model B should be extended with the spatial-temporal evolution of the fireball, which is an ongoing efforts~\cite{Kitazawa:2020kvc}.

\underline{\textbf{Non-equilibrium chiral fluid dynamics (N$\chi$FD)}} is a dynamical model that
couples the chiral condensate with the evolving fluid~\cite{Nahrgang:2011mg,Nahrgang:2011mv,Nahrgang:2011vn,Herold:2014zoa,
Herold:2016uvv,Herold:2018ptm,Paech:2003fe}. The chiral condensate $\sigma$, treated as the critical mode, evolves according to the Langevin equation with the effective potential obtained from the effective theory ({\it e.g.,} (P)QM model \cite{Sasaki:2011sd}, linear sigma model \cite{GellMann:1960np,Scavenius:2000qd}). It also provides source terms for the hydrodynamic equations for the heat bath evolution of the fluid. In~\cite{Herold:2014zoa,Herold:2016uvv,Herold:2018ptm},  N$\chi$FD  has been numerical implemented  to study various dynamical effects near the QCD phase transition. It was found  that variance and kurtosis for the net-proton multiplicities  are enhanced near the critical point compared with the values with the crossover phase transition.

\underline{\textbf{Fluctuating Hydrodynamics}} extends traditional dissipative hydrodynamics with stochastic fluctuations according to the fluctuation-dissipation theorem~\cite{Kapusta:2011gt}, where fluctuations are treated as white noise. Fluctuating Hydrodynamics was first proposed in a non-relativistic form~\cite{Landau:1959,Lifshitz:1980} and extended to the relativistic case in~\cite{Kapusta:2011gt}. In principle, by identifying the slow mode associated with critical fluctuations and properly implementing the EoS and transport coefficients, one could study the dynamics of critical fluctuations and their influence on experimental observables using the fluctuating hydrodynamics framework. For (0+1)D boost invariant expanding background, analytical calculations showed that the magnitude of the correlation function enhances due to the increased thermal conductivity near the QCD critical point~\cite{Kapusta:2012zb}. In practice, especially for numerical implementations, one should carefully deal with the {multiplicative} noise~\cite{Murase:2013tma,Kovtun:2014hpa,Arnold:1999va} and the dependence of the grid sizes, especially for the case near the phase transition. The renormalization of the equation of state, the transport coefficients~\cite{Kovtun:2011np,Chafin:2012eq}, and the proper treatment for the numerical simulations are still challenging and under development~\cite{Murase:2016rhl,Hirano:2018diu,Nahrgang:2017oqp,Bluhm:2018plm,Singh:2018dpk}.

\underline{\textbf{ {Hydrodynamics-kinetics}}} evolves stochastic fluctuations with the deterministic kinetic equation for two-point correlation function, while treating the background evolution hydrodynamically~\cite{Akamatsu:2016llw,An:2019osr,Akamatsu:2017rdu,Akamatsu:2018vjr,Martinez:2018wia}. It introduces renormalized transport coefficients and the equation of state, which naturally absorbs the lattice spacing dependence in the numerical implementation of fluctuating hydrodynamics mentioned above. The hydrodynamics-kinetic approach could reproduce the long-time tails of the hydrodynamics fluctuations~\cite{Kovtun:2012rj,Chafin:2012eq,Martinez:2017jjf,Martinez:2018wia} and estimate the relevant scale of critical fluctuations near the critical point~\cite{Akamatsu:2018vjr}. Besides the two-point function, the kinetic equation for three-point and four-point functions has also been developed recently~\cite{Pratt:2019fbj, An:2020vri}.

\underline{\textbf{Hydro+}} encodes the critical fluctuations in the kinetic equation of the slow modes and couples it with the hydrodynamic evolution~\cite{Stephanov:2017ghc}. Here, the slowest mode $\phi_{\bm{Q}}(t,\bm{x})$ is identified as the Wigner transformed two-point function of entropy per baryon, which is driven to out-of-equilibrium near the critical point. With the generalized EoS, shear viscosity and bulk viscosity are modified by the slow mode $\phi_{\bm{Q}}(t,\bm{x})$, dynamics of critical fluctuations naturedly couples with the evolving fluid background. Hydro+ has been numerically implemented in a simplified system for a rapidity-independent fireball undergoing radial flow~\cite{Rajagopal:2019xwg}, and ideal Gubser flow \cite{Du:2020bxp}. The dynamics of the slow modes $\phi_{Q}$ have negligible effects on the fluid evolution. The out-of-equilibrium corrections to generalized entropy is within 0.1\% in the numerical simulations. Besides the dynamics of the two-point function for the slowest modes, the extension of the hydro+ by involving other slow modes, called hydro++, are also under development~\cite{An:2019csj, An:2020jjk}.

\underline{\textbf{Dynamical universal scaling}} comes from the competition between the relaxation rate of the slow mode near the critical point and the expanding rate of the evolving system. Due to the critical slowing down effects, the critical modes will be driven out of equilibrium, which lead to correlated regions with characteristic length scales, called Kibble-Zurek scales, defined by the correlation length at which these two rates equal to each other. It was realized that, within the framework of Kibble-Zurek mechanism (KZM), one could construct some universal variables near the critical point that are independent of some non-universal factors, such as the evolving trajectory of the system approaching the critical point~\cite{Francuz:2015zva,Nikoghosyan:2013fqa}. Recently, the universal behavior of non-conversed order parameter field and conserved quantities near the QCD critical point has been studied within the framework of model A and model B~\cite{Mukherjee:2016kyu,Wu:2018twy,Wu:2019qfz,Akamatsu:2018vjr}. It was found that the oscillating behavior for the higher cumulants of net protons can be drastically suppressed, which converge into approximate universal curves with these constructed Kibble-Zurek functions~\cite{Wu:2018twy}.

\section{Softening and clustering across the first-order phase boundary}

For the system evolving across the first-order phase boundary, the bulk matter tends to separate into the two coexisting phases with the development of instabilities. According to~\cite{Gibbs}, such processes in the coexistence region are classified into two categories: nucleation and spinodal decomposition.

\underline{\textbf{Nucleation}} belongs to nonlinear instability, which requires the formation of a sufficient large nucleus of the stable phase to overcome the barrier of free energy between stable and metastable minimums. The classical nucleation theory has been developed many years ago~\cite{Becker}, together with the following extensions~\cite{Lothe,Cahn, Langer:1967ax, Langer:1969bc, Zeng:1991}. The essential point is the nucleation rate that describes the decay probability per time and per volume of a metastable state, $\Gamma \sim e^{-\Delta\mathcal{F}_c}$,
is mainly determined by the cost of the free energy $\Delta\mathcal{F}_c=\Delta \mathcal{F}(R_c)$ (where $R_c$ is the critical size of the bubble formed in a system). In heavy-ion physics, the nucleation rate of hadronic bubbles formed in the QGP is estimated through extending the classical nucleation theory~\cite{Csernai:1992tj,Csernai:1992as,Csernai:1992bs,Zabrodin:1998dk,Mishustin:1998eq,Alamoudi:1999ti,
Shukla:2000dx,Shukla:2001xv,Bessa:2008nw}, in which the supercooling (or reheating) and the probability of bubble formation are influenced by many factors, such as the barrier height between two minimums of free energy, viscosity, and the expansion scenario. The effects of the supercooling or reheating have recently been observed in dynamical model simulations with the first-order phase transition, such as model A based on Langevin dynamics~\cite{Jiang:2017mji} or N$\chi$FD~\cite{Nahrgang:2011mv,Nahrgang:2011vn,Paech:2003fe,Herold:2014zoa}. In the rapidly expanding fireball, such systems could further evolve into the unstable region where the spinodal instabilities become dominant, which will be addressed follow.

\underline{\textbf{Spinodal decomposition}} belongs to linear instability with small fluctuation of the order parameter field growing instantly and uniformly throughout the volume when the system evolves into the negative curvature region of the free energy. The basic theory of spinodal decomposition is the Cahn-Hilliard equation~\cite{Cahn}, which has been widely studied in metallurgy. In heavy-ion physics, the spinodal decomposition has been investigated within the framework dissipative hydrodynamics using an approximate equation of state for the coexistence region~\cite{Randrup:2010ax,Randrup:2009gp,Steinheimer:2012gc,Steinheimer:2013xxa,Steinheimer:2013gla}.
In this model, the nuclear spinodal separation between confined and deconfined phases was observed, and its experimental consequences, such as the increasing density moments arising from the spinodal amplification of spatial irregularities, have also been studied.

Although some progress has been made during the past years (Please also refer to Refs~\cite{Skokov:2009yu, Pratt:2017lce}), the dynamics near the first-order phase transition is still poorly understood due to the complexities of the evolving system. For instance, systems at the unstable region are easily driven out of equilibrium, which requires an additional study on the validity of hydrodynamics near the first-order phase transition region. More attention and effort from the theoretical and phenomenological side are still needed to study and predict the related experimental observable.

\section{Experimental observables at RHIC BES}

This section will briefly review the experimental observable measured at the RHIC BES program phase I, primarily focusing on soft observables to probe the bulk properties of the QGP and observables to probe the critical point and the first-order phase transition.

\subsection{Soft observables to probe the bulk properties of QGP}

\underline{\textbf{Particle yields, $p_T$ spectra, and flow }} in Au+Au collisions at the RHIC BES program are important soft observables to probe the QGP properties and the QCD phase structure. More specifically, the $p_T$-integrated yield at mid-rapidity provides detailed information on hadronic chemistry, from which the chemical freeze-out temperature and chemical potentials can be extracted using the statistical hadronization model~\cite{Andronic:2017pug,Andronic:2008gu,
Adamczyk:2017iwn}. The mean $p_T$ of identified particles are directly proportional to the amount of radial flow of the evolving system, which helps to elucidate the size of pressure gradients in the initial profiles, the speed of sound, and bulk viscosity of the evolving systems~\cite{Ryu:2015vwa,Song:2009rh}. Various flow observables are sensitive to the transport properties and initial state fluctuation of the QGP fireball~\cite{Gale:2012rq,Qiu:2012uy,Heinz:2013bua,Teaney:2013dta,
Zhu:2016puf,Pang:2015zrq,Zhao:2017yhj,Zhao:2017rgg,Moravcova:2020wnf,Giacalone:2020byk}. A comparison of the $\sqrt{s}$ dependence of the elliptic flow with hydrodynamics + transport hybrid models showed that the effective shear viscosity increases at low collision energies \cite{Karpenko:2015xea}. The work~\cite{Shen:2020jwv} further demonstrated that the elliptic flow measurements as a function of collision energy and rapidity could set strong constraints on the $T$ and $\mu_B$ dependence of the QGP specific shear viscosity. Recently, the STAR collaboration found that the particle multiplicity scaled triangular flow $v_3$ as a function of collision energy exhibited a minimum at $\sqrt{s_\mathrm{NN}} \sim 20$\,GeV \cite{Adamczyk:2016exq}, which hinted at a softening of the equation of state around the minimum. However, this non-monotonic behavior can be reproduced by the event-by-event hybrid framework without a critical point \cite{Shen:2020gef}. Therefore, it is essential to understand the interplay among the duration of dynamical initialization, the variation of the speed of sound, and the $T$ and $\mu_B$ dependence of the specific shear viscosity.

\underline{\textbf{The net proton rapidity distributions}} have been measured in a few experiments during the past years~\cite{E-802:1998xum,Ahle:1999in,Barrette:1999ry,Arsene:2009aa,Anticic:2010mp}. These measurements can elucidate the initial baryon stopping and the dynamics of conserved charge density currents inside the QGP. The latter is controlled by the medium's charge diffusion constants and heat conductivity. The baryon diffusion during the hydrodynamic phase evolves baryon charges from forwarding rapidity back to the central rapidity region~\cite{Denicol:2018wdp, Li:2018fow, Du:2019obx, Wu:2021ypv}. These dynamics are driven by the inward-pointing spatial gradients of $\mu_B/T$, which act against local pressure gradients. Because the net proton rapidity distribution is sensitive to both the initial state stopping and baryon diffusion \cite{Denicol:2018wdp}, independent experimental observables are needed to disentangle these two effects. Recently, charge balance functions were proposed to independently constrain the charge diffusion constants \cite{Pratt:2019pnd, Pratt:2021xvg}. A larger diffusion constant in the QGP medium leads to wider azimuthal distribution for the $K^+K^-$ and $p\bar{p}$ correlations.

\subsection{Observables to probe the critical point and the first-order phase transition}

\underline{\textbf{Multiplicity fluctuations of net protons}} serves as a proxy for the net-baryon fluctuations, which is regarded as a promising experimental observable for searching the QCD critical point~\cite{Hatta:2003wn,Kitazawa:2012at,Kitazawa:2011wh}. The skewness and kurtosis of net proton distribution $S\sigma$ and $\kappa\sigma^2$, with the momentum coverage $|y|<0.5$ and $0.4<p_T<2$ GeV, have been systematically measured in Au+Au collisions from 7.7 to 200 A GeV~\cite{Adam:2020unf,Abdallah:2021fzj}, which shows potential features that may hint the critical point and first-order phase transition. With collision energy decreases, the net proton $\kappa\sigma^2$ first decreases and then increases with a minimum at $\sqrt{s_\mathrm{NN}} \sim 20$\,GeV. However, the error bars in the RHIC BES I measurements below 19.6 GeV remain too large to confirm any non-monotonic $\sqrt{s}$ dependence.




\underline{\textbf{Light nuclei productions}} are argued to sensitive to the relative density fluctuation at the freeze-out surface, which is proposed as a sensitive observable to probe the QCD critical point and the first-order phase transition~\cite{Sun:2017xrx,Sun:2018jhg}. Recently, the STAR Collaboration has systematically measured the yields of light nuclei at the RHIC BES program and found that the coalescence parameters of (anti-)deuteron and the yield ratio of light nuclei, $N_t\cdot N_p/N_d^2$ show a non-monotonic energy dependence with a dip and a peak around $\sqrt{s_\mathrm{NN}}=20$\,GeV in central Au+Au collisions, respectively~\cite{Adamczyk:2016gfs,Adam:2019wnb,Zhang:2019wun}. Dynamical models without critical point or first-order phase transition fail to reproduce the non-monotonic behavior of light nuclei ratio ~\cite{Liu:2019nii,Sun:2020uoj,Deng:2020zxo,Zhao:2020irc}. It was recently found that the attractive and repulsive interaction between nucleons near the critical point was found to play an important role in the clustering of the light nuclei \cite{Shuryak:2018lgd,Shuryak:2019ikv,Shuryak:2020yrs,DeMartini:2020hka,DeMartini:2020anq}. Ref.~\cite{Sun:2020pjz} also study the light nuclei production near first-order phase transition within the framework of transport model that includes a first-order chiral phase transition and found the enhanced yield ratio near the transition region.
Last but not least, a recent work \cite{Oliinychenko:2020znl} pointed out that the weak decay corrections to the proton spectrum need careful attention as it could potentially contaminated the collision energy dependence of the measured $N_t\cdot N_p/N_d^2$ ratio.

\section{Outlook}

Over the past two decades, the collective advancements in developing quantitative theoretical frameworks for relativistic nuclear collisions and high-quality flow measurements at top RHIC and the LHC energies have been driving our field to a precision era. With the increasing complexity in theoretical models, statistics and data science techniques such as Bayesian Inference have become a standard tool, which systematically constrains the QGP transport properties and initial-state fluctuations from various available soft observables~\cite{Everett:2020yty, Everett:2020xug,Moreland:2018gsh,Bernhard:2019bmu}.

The RHIC Beam Energy Scan program has excited a wave of theoretical developments on dynamical modeling with full (3+1)D simulations that go beyond Bjorken's boost-invariant assumption. With dynamical initialization schemes which interweave the initial 3D collision dynamics with hydrodynamic simulations, we are starting to quantify initial baryon stopping and study the collectivity of the QGP in a baryon-rich environment. Confronting flow measurements from the upcoming RHIC BES program phase II and future FAIR/NICA experiments, we can elucidate how the QGP transport properties change as a function of net baryon density. We also expect that the Bayesian analysis with the BES phase II measurements at RHIC will quantitatively characterize the QGP viscosity and charge diffusion constant.

In the meantime, experiment measurements on the multiplicity fluctuations of net proton and light nuclei productions at the RHIC BES program, together with the model studies, hint for the possible existence of the critical point and the first-order phase transition boundary. Quantitative description of these measurements requires more realistic and sophisticated model development on the theoretical side. Especially, the phenomenological model with the critical slowing down and first-order phase transition effects, together with a proper equation of state for the hot QCD system, needs to be developed consistently with the expanding collective background for such short-lived and tiny fireballs created at heavy-ion collisions. In parallel, high statistic measurements from the coming RHIC BES phase II could provide more confidence on the potential critical point signals.  Furthermore, the larger rapidity acceptance and collisions at lower energies through fixed target setups in the RHIC BES II allowsfor detecting a broader range of the QCD phase diagram than that in the phase I program.
Together with the description from sophisticated dynamical models, our community will provide deeper insights into the QCD phase structure. \\

\textbf{Acknowledgments}

S.W. and H.S. is supported by the NSFC under grant No. 12075007 and No. 11675004.
S.W. is also supported by the China Postdoctoral Science Foundation under Grant No. 2020M680184 and NSFC under grant No. 11947236.  C.S. is supported by the U.S. Department of Energy under grant number DE-SC0013460 and the National Science Foundation under grant number PHY-2012922.
This work is also supported in part by the U.S. Department of Energy, Office of Science, Office of Nuclear Physics, within the framework of the Beam Energy Scan Theory (BEST) Topical Collaboration.




\bibliographystyle{apsrev}
\bibliography{CPL}
\end{multicols}
\end{document}